\documentclass[12pt,leqno]{amsart}
\usepackage{amsmath,amssymb,amscd,amsthm}
\usepackage[dvips]{graphics,epsfig}
\usepackage{graphicx}
\usepackage{subfigure}
  \usepackage{paralist}
  \usepackage{graphics} 
  \usepackage{epsfig} 
\usepackage{graphicx}  \usepackage{epstopdf}
 \usepackage[colorlinks=true]{hyperref}
\hypersetup{urlcolor=blue, citecolor=red}

\newtheorem{theorem}{Theorem}
\newtheorem{lemma}{Lemma}
\newtheorem{proposition}{Proposition}

\newtheorem{definition}{Definition}
\newtheorem{corollary}{Corollary}

\newtheorem{claim}{Claim}

\newcommand{\f}[2]{\frac{#1}{#2}}
\newcommand{\dpr}[2]{\left\langle #1,#2 \right\rangle}

\newcommand{\be}{\beta}

\newcommand{\de}{\delta}

\newcommand{\De}{\Delta}

\newcommand{\ka}{\kappa}
\newcommand{\la}{\lambda}

\newcommand{\si}{\sigma}

\newcommand{\vp}{\varphi}
\newcommand{\om}{\omega}

\newcommand{\rd}{{\mathbf R}^d}

\newcommand{\rone}{\mathbf R^1}

\newcommand{\ch}{\mathcal H}

\newcommand{\bv}{\mathbf v}

\newcommand{\p}{\partial}

\newcommand{\beq}{\begin{equation}}
\newcommand{\eeq}{\end{equation}}
\newcommand{\beqna}{\begin{eqnarray*}}
\newcommand{\eeqna}{\end{eqnarray*}}
\newcommand{\beqn}{\begin{equation*}}
\newcommand{\eeqn}{\end{equation*}}
\newcommand{\bp}{\begin{proof}}
\newcommand{\ep}{\end{proof}}
\newcommand{\bprop}{\begin{proposition}}
\newcommand{\eprop}{\end{proposition}}
\newcommand{\bt}{\begin{theorem}}
\newcommand{\et}{\end{theorem}}
\newcommand{\bex}{\begin{Example}}
\newcommand{\eex}{\end{Example}}
\newcommand{\bc}{\begin{corollary}}
\newcommand{\ec}{\end{corollary}}
\newcommand{\bcl}{\begin{claim}}
\newcommand{\ecl}{\end{claim}}
\newcommand{\bl}{\begin{lemma}}
\newcommand{\el}{\end{lemma}}

\title
[Spectral stability of Klein-Gordon standing waves]
{Spectral stability analysis for standing waves of a perturbed Klein-Gordon equation}

\author [ A. Demirkaya and P.G. Kevrekidis and M. Stanislavova and A. Stefanov]{}



\email{demirkaya@hartford.edu}
\email{kevrekid@math.umass.edu}
\email{stanis@math.ku.edu}
\email{stefanov@ku.edu}

\thanks{{Kevrekidis is supported by NSF-DMS-1312856,
the US-AFOSR under grant FA9550-12-10332, from FP7-People under grant IRSES-606096 and the Binational (US-Israel) Science Foundation through grant
2010239. Stanislavova  supported in part by  
  NSF-DMS \# 1211315. 
Stefanov supported in part by  
NSF-DMS \# 1313107.}}

\begin{document}
\maketitle

\centerline{\scshape Aslihan\ Demirkaya}
\medskip
{\footnotesize
 \centerline{Department of Mathematics}
  \centerline{University of Hartford}
 \centerline{200 Bloomfield Avenue, West Hartford, CT 06117, USA}

}
\medskip

\centerline{\scshape Panayotis G.\ Kevrekidis}
\medskip
{\footnotesize
 \centerline{Department of Mathematics and Statistics}
  \centerline{University of Massachusetts}
 \centerline{Amherst, MA 01003-4515, USA}

}
\medskip

\centerline{\scshape Milena\ Stanislavova}
\medskip
{\footnotesize
 \centerline{Department of Mathematics}
  \centerline{University of Kansas}
 \centerline{1460 Jayhawk Boulevard,  Lawrence KS 66045--7523}

}
\medskip

\centerline{\scshape Atanas\ Stefanov}
\medskip
{\footnotesize
 \centerline{Department of Mathematics}
  \centerline{University of Kansas}
 \centerline{1460 Jayhawk Boulevard,  Lawrence KS 66045--7523}

}
\medskip


\begin{abstract}
In the present work, we introduce a new 
$\mathcal{PT}$-symmetric variant of the Klein-Gordon
field theoretic problem. We identify the standing
wave solutions of the proposed class of equations
and analyze their stability. In particular, we
obtain an explicit frequency condition, somewhat
reminiscent of the classical Vakhitov-Kolokolov
criterion, which sharply separates the regimes
of spectral stability and instability. Our
numerical computations corroborate the relevant
theoretical result.
\end{abstract}

\section{Introduction} \label{intro}

Over the past 15 years, the original proposal of Bender and
co-workers that systems with $\mathcal{PT}$-symmetry may constitute
relevant extensions of the usual Hermitian quantum mechanical
models has gained considerable traction. Part of the reason for
the interest in this theme has been the theoretical 
proposal~\cite{ziad,Ramezani,Muga}, but perhaps especially so
the experimental implementation~\cite{salamo,dncnat} in linear and nonlinear
optics of systems that follow the proposed $\mathcal{PT}$-symmetric
dynamics. More recently, similar systems have been implemented
in electrical~\cite{R21,tsampas2} and mechanical~\cite{bend_mech}
linear systems, as well in the realm of
whispering-gallery microcavities~\cite{pt_whisper} and in a 
$\mathcal{PT}$-symmetric dimer of Van-der-Pol oscillators in~\cite{bend_kott}.

In numerous ones among these systems 
(e.g. in~\cite{R21,tsampas2,bend_mech,pt_whisper,bend_kott}), the
underlying linear dynamics is of the oscillator type i.e., it involves
a dimer of two oscillators, one with loss and one with gain, typically
in the form of a linear dashpot. The relevant oscillator pair reads
(at the linear level):
\begin{eqnarray}
\ddot{u}=-\omega^2 u - \epsilon v - \gamma \dot{u}
\label{eqn0a}
\\
\ddot{v}=-\omega^2 v - \epsilon u + \gamma \dot{v}
\label{eqn0b}
\end{eqnarray}
where $\omega$ represents the frequency of the oscillators, $\epsilon$
their coupling, while $\gamma$ is strength of the loss/gain in the
two oscillators $u$ and $v$. 
This, in turn, has motivated
a number of studies that considered both the discrete~\cite{asli1},
as well as the long-wavelength continuum~\cite{asli2,pgk} 
generalization of such models in the realm of Klein-Gordon
partial differential equations of the form (now for the field $u(x,t)$):
\begin{eqnarray}
u_{tt}=u_{xx} + g(u) + W(x) u_t
\label{eqn0}
\end{eqnarray}
where $W(x)$ is anti-symmetric in order to enable regions of
loss (with $W(x)<0$) and gain (with $W(x)>0$); $g(u)$ contains
the nonlinearity potentially present in the model.

Here, we explore the long-wavelength limit of a modified form
of the oscillator problem whereby the oscillation of $u$
involves a dashpot effect from $v$ and that of $v$ a
gain effect from $u$. 
This type of velocity dependent
coupling has been argued, for instance, to exist
in the coupling of pendula in the recent experiments and
associated modeling of~\cite{anderson}. Our own 
investigation, however, is chiefly motivated by the 
the continuum properties of the corresponding 
long-wavelength limit mathematical system which in this case will be
of the following Klein-Gordon form, again for the field $u(x,t)$:
\begin{equation}
\label{10}
u_{tt}+ i \be W(x) u_t- \De u +u - f(|u|^2) u=0, \ \  (t,x)\in \rone_+\times \rd
\end{equation}
where $W$ is a real-valued and bounded potential   and $\be$ is a real parameter.  
We will give general stability conditions involving $W$,
although our interest will be chiefly towards the $\mathcal{PT}$-symmetric
case, whereby the invariance under $x \rightarrow -x$ and 
$t \rightarrow -t$, as well as $i \rightarrow -i$ clearly indicates
that $W(-x)=W(x)$ and $W$ should be an even function to ensure 
$\mathcal{PT}$-symmetry. An alternative way of thinking of this 
partial differential equation (PDE) is as a Schr{\"o}dinger 
model with an added inertial term. This, in turn, suggests
the underlying conservative nature of this PDE, which we will not
explore further below but which will be somewhat implicit in our
spectral considerations.

In what follows, we will be interested in standing wave solutions in the form $e^{i \om t}\vp(x)$, with real-valued carrier $\vp$, which naturally satisfy 
\begin{equation}
\label{20}
-\De \vp +(1-\om^2) \vp-\be \om W \vp-f(\vp^2)\vp=0, \ \  x\in \rd.
\end{equation}
The equation \eqref{20} will have homoclinic orbit, pulse-like 
solutions under appropriate conditions on the nonlinearity $f$ and the
function $W(x)$. For example, if $f(z)=z^{(p-1)/2}$ for some  $p>1$ and $W$ is a bounded function, one can show that a (positive) solution to \eqref{20} may be obtained as a (multiple of) the solution to the following constrained minimization problem 
$$
\left\{
\begin{array}{l}
\int_{\rd} [ |\nabla h(x)|^2+(1-\om^2) h^2(x)-\be \om W(x) h^2(x)]dx\to \min  \\
\int_{\rd} h^{p+1}(x) dx = 1. 
\end{array}
\right. 
$$
In all these cases, the present work will focus on the spectral stability of such solutions. In order to study this question, we linearize \eqref{10} by $u=e^{i \om t}[\vp+ v(t,x)]$.  By keeping the linear terms and 
 ignoring all higher terms $O(v^2)$, we arrive at 
\begin{equation}
\label{30}
v_{tt}+i(2\om + \be  W) v_t+[-\De v+(1-\om^2)v -\be \om W v - f(\vp^2)v -2\vp^2 f'(\vp^2)\Re v]=0.
\end{equation}
Furthermore, introducing the vector $\bv$ 
\begin{equation}
\label{40}
\bv_{tt}+ J \bv_t+\ch \bv=0,
\end{equation}
where 
\begin{eqnarray*}
J = J(\be) &=& \left(\begin{array}{cc}
0 & -(2\om+\be W) \\
(2\om +\be W) & 0 
\end{array}
\right), \ch=\left(\begin{array}{cc}
L_+ & 0 \\
0 & L_- 
\end{array}
\right) \\
L_+ =L_+(\be) &=& -\De +(1-\om^2) -\be \om W  - f(\vp^2) -2\vp^2 f'(\vp^2)\\
L_- =L_-(\be)&=&  -\De +(1-\om^2) -\be \om W  - f(\vp^2).
\end{eqnarray*}
We now give a definition for stability/instability of such linearizations. 
\begin{definition}
We say that the linearized problem \eqref{40} is spectrally unstable, if there is $\la: \Re \la>0$ and $\Psi\in D(\ch)$, so that 
\begin{equation}
\label{50}
\la^2 \Psi+\la J \Psi+ \ch \Psi=0.
\end{equation}
Otherwise, we say that the linearized problem \eqref{40} is spectrally stable. 
\end{definition}
We should mention that eigenvalue problems of this type have been considered in the literature; see e.g.~\cite{SS} and references therein. In fact, 
they are frequently referred to as operator pencils. This is due to the 
quadratic dependence on the eigenvalue parameter in \eqref{50}. The 
following general result helps us decide about the stability of such pencils. 

\begin{theorem} \cite{SS}
\label{theo:10}
Assume that the operators $J, \ch$ satisfy the following assumptions: 
\begin{enumerate}
\item[(A)] $L^2=X^+\oplus X^-$, so that  
$$\overline{\ch u}=\ch \bar{u}, \ \  \ch:X^\pm\cap D(\ch) \to X^\pm, \ \ 
   \ch^*=\ch
$$
\item[(B)]
$$
 \overline{J u}=J \bar{u},\ \   J:X^\pm\to X^\mp, \ \  J^*=-J, 
  \ \   \forall \tau>>1: J(H+\tau)^{-1}\in B(L^2)
$$

\item[(C)]
$$
\left\{
\begin{array}{l}
\ch\phi=-\de^2 \phi,  \ch|_{\{\phi\}^\perp}\geq 0; \si_{a.c.}(\ch)\subset [\ka^2, \infty), \ka>0 \\
Ker[\ch]=span[\psi_0,  \ldots, \psi_l],  \|\psi_j\|=1, j=0, \ldots, l\\
\psi_0\in X^-; \{\phi, \psi_1, \ldots, \psi_l\}\in X^+; \dpr{\psi_i}{\psi_j}=0, j\neq k;  
\end{array}
\right.
$$
\item[(D)]
$$\dpr{\psi_j}{J\psi_0}=0, \ j=1, \ldots, n.$$

\end{enumerate}
Then, 
\begin{itemize}
\item the pencil \eqref{50}  is spectrally unstable if 
$$
\dpr{\ch^{-1}[J\psi_0]}{J\psi_0}>-1
$$
 
\item the pencil \eqref{50}  is spectrally
  stable, if 
$$
\dpr{\ch^{-1}[J\psi_0]}{J\psi_0}\leq  -1
$$
 
 \end{itemize}
 {\bf Note: } For the parameter $l\geq 0$,   $l=0$ is allowed. That is, the kernel of $\ch$ may  be one dimensional, in which case, $(D)$ is vacuous. 

\end{theorem}

Our aim in the present work is to quantify this general theorem
in the special case of the operator pencils discussed above in~(\ref{40}).
More specifically, in Section 2, we give the precise condition that
is relevant to our operator pencil. In section 3, we give a series
of specific examples for power law nonlinearities and particular
forms of $W(x)$. In section 4, we consider some typical ones among
these examples numerically and corroborate the prediction of the
theorem. Finally, in section 5, we present our conclusions and
propose some possibilities for future work.

\section{Main results}
 Let us start by saying a few words on how our principal example 
of \eqref{40}  satisfies the requirements of Theorem \ref{theo:10}.  First, if $W$ is appropriately decaying\footnote{We will also consider the example $W(x)=1$ in which case this will change, but we discuss this separately.}, the essential spectrum of $\ch$ is $[1-\om^2, \infty)$ by Weyl's theorem. 
The split $L^2=X^+\oplus X^-$ is then not hard to guess, namely  
$
X^+=\left\{ \left(\begin{array}{c} f \\ 0 \end{array}\right): f\in L^2(\rone)\right\}$, 
$
X^-=\left\{ \left(\begin{array}{c} 0 \\ g \end{array}\right): g\in L^2(\rone)\right\}. 
$
Clearly, the requirements $(A), (B)$ are satisfied, since the operators $L_\pm$ are self-adjoint with a domain $H^2(\rone)$ and $J: X^\pm\to X^\mp$, while $\ch: X^\pm\cap H^2 \to X^\pm$. 
If the wave $\vp>0$ (as is the case for the prototypical ground-state
pulse that will interest us herein), we have that $L_- [\vp]=0$. Thus, in the 1 D case by Sturm-Liouville theory, it follows that $L_-\geq 0$, with a (normalized)  eigenfunction $\psi_0:= \|\vp\|^{-1} \vp$. 
The requirement that $L_+$ satisfies the condition $(C)$ is non-trivial. Namely, we need $L_+$ to have at most one negative eigenvalue (counting multiplicities)  and $Ker[L_+]$ be empty or $Ker[L_+]=span[\psi_1, \ldots, \psi_l]$.  If the second possibility occurs, then we need to make sure, by $(D)$,  that $\dpr{\psi_j}{J\vp}=0$. This will actually turn out to be automatic in our case. 

The following is the main result of the present contribution. 
\begin{theorem}
\label{theo:20}
Let   $\om \in (-1,1)$  and assume the problem \eqref{20} has a positive smooth solution (in both $x$ and $\om$ variables) $\vp_\om(x)$,  
$\lim_{|x|\to \infty} \vp_\om(x)=0$.  Assume  
\begin{equation}
\label{cond:1}
 n(L_+)= \# \{\la\in \si(\ch): \la<0\} =  1. 
\end{equation}
Next assume that either $d=1$ or if $d\geq 2$, then $Ker[L_-]=span[\vp]$

Then, the wave $\vp_\om$ is spectrally stable if and only if 
\begin{equation}
\label{80}
\p_\om[  \int (2\om+\be W)\vp_\om^2(x)  dx] \leq 0.
\end{equation}

\end{theorem}
\noindent {\bf Remark:} For  $f(z)=z^{\f{p-1}{2}}$, we have that $L_+\leq L_-$, and in fact $\dpr{L_+ \vp}{\vp}=-(p-1) \int \vp^2(x) dx<0$, so  $L_+$ always has negative point spectrum. The requirement \eqref{cond:1} is therefore asking for such spectrum to be reduced to a single point. 
\begin{proof}
We first show that it suffices to require only \eqref{cond:1} and then our pencil \eqref{40} satisfies the assumptions of Theorem \ref{theo:10}.  
Next, let us discuss the conditions on $Ker[L_+]$. If $Ker[L_+]=\emptyset$, there is nothing else to do, this is the case $l=0$ in Theorem \ref{theo:10}. If however $Ker[L_+]=span[\psi_1, \ldots, \psi_l]$, we need to have that $\dpr{\psi_j}{J \psi_0}=0$. That is, we need to have $(2\om +\be W)\vp \perp Ker[L_+]$. To prove this, take the  equation \eqref{20} defining our standing wave 
and take a derivative with respect to $\om$. We obtain 
$$
L_+[\p_\om \vp]-2\om\vp-\be W \vp=0. 
$$
It follows that 
$
L_+[\p_\om \vp]=(2\om +\be W) \vp.
$, whence $\p_\om \vp=L_+^{-1}[(2\om +\be W) \vp]$.

Taking dot product of the last identity with $\psi_j, j=1, \ldots, l$, we get $\dpr{\psi_j}{(2\om +\be W) \vp}=0$, which is the condition $(D)$ in Theorem \ref{theo:10}. 

It now remains to compute the quantity $\dpr{\ch^{-1}(J \psi_0)}{J \psi_0}$. We have 
\begin{eqnarray*}
\ch^{-1} J\psi_0= -\f{1}{\|\vp\|} \ch^{-1} \left(\begin{array}{c} (2\om+\be W) \vp \\ 0\end{array} \right) = 
 -\f{1}{\|\vp\|}  L_+^{-1}[(2\om+\be W) \vp],
\end{eqnarray*}
whence 
\begin{eqnarray*}
& & \dpr{\ch^{-1} J\psi_0}{J\psi_0}=\f{1}{\|\vp\|^2}\dpr{ L_+^{-1}[(2\om+\be W) \vp]}{(2\om+\be W) \vp}=\\
&= & \f{1}{\|\vp\|^2} \dpr{\p_\om \vp}{(2\om+\be W) \vp}= \f{\p_\om[\int \om \vp^2(x) +\f{ \be W(x) }{2} \vp^2(x) dx]-\|\vp\|^2}{\|\vp\|^2}. 
\end{eqnarray*} 
Thus, for stability it is necessary and sufficient to have that the last expression is less than $-1$, 
 so we arrive at   \eqref{80}.  

\end{proof}

It is also relevant to note here that this condition appears to be a
natural generalization for the present setup of the famous Vakhitov-Kolokolov
condition~\cite{vakhitov} for the stability of ground state solitary waves of
the nonlinear Schr{\"o}dinger equation.

\section{Examples}
In this section, we consider several examples which fall within the framework of Theorem \ref{theo:20}. 
\subsection{The case $W(x)=1$} 
Our first example is for $f(z)=z^{\f{p-1}{2}}$, $p>1$ and the potential is a constant function, that is $W(x)=1$. The dimension $d\geq 1$ is arbitrary. 
Fix $\be$. In the case of $d=1$, \eqref{20} becomes 
\begin{equation}
\label{100}
-\vp''+(1-\om^2-\be \om) \vp - \vp^{p}=0.
\end{equation}
Note that in this case, we actually have solutions in the set $\{\om: 1-\om^2-\be \om>0\}$.  The same condition is required for the spectral gap condition to hold, so we assume it. That is 
\begin{equation}
\label{110}
\f{-\be-\sqrt{\be^2+4}}{2}<\om< \f{-\be+\sqrt{\be^2+4}}{2}
\end{equation} 
 All positive  solutions to \eqref{100} are then in the form 
\begin{equation}
\label{120}
\vp_\om(x)=(1-\om^2-\be \om)^{\f{1}{p-1}} \phi_{d,p}(x\sqrt{1-\om^2-\be \om} ).
\end{equation} 
where $\phi$ is a fixed function depending on $d,p$ only. This is the uniqueness result of~\cite{Kwong}. 
Next, the operator $L_+$ has one simple negative eigenvalue, while $L_-$ has only $\vp$ in its kernel, and in fact $Ker[L_+]=span[\p_1 \vp, \ldots, \p_d \vp]$ -  this was  shown in a series of papers by Weinstein, \cite{Wei1}, Shatah, \cite{Sh100} and   Kwong, \cite{Kwong}. It remains to compute the quantity in 
\eqref{80} and to solve the inequality. We have 
\begin{eqnarray*}
\p_\om[ (2\om+\be) \|\vp_\om\|^2 ]=(1-\om^2-\be \om)^{\f{2}{p-1}-\f{d}{2}}
\left(2-\f{(2\om+\be)^2(\f{2}{p-1}-\f{d}{2})}{1-\om^2-\be \om)}\right).
\end{eqnarray*}
Setting $c_p:= \f{2}{p-1}-\f{d}{2}$, we can rewrite the stability condition $\p_\om[ (2\om+\be) \|\vp_\om\|^2 ]\leq 0$ as follows 
$$
2\leq c_p \f{(2\om+\be)^2}{(1-\om^2-\be \om)}. 
$$
Clearly, if $c_p\leq 0$ (corresponding to   $p\geq 1+\f{4}{d}$), the inequality fails leading to instability. 
Otherwise, we reduce to a quadratic  inequality for $\om$, which has the solutions 
$$
-\f{\be}{2} - 
\f{1}{2} \sqrt{\f{\be^2+4}{2c_p+1}} \leq \om\ \  \  \textup{and} \ \    \ \ \ \om\geq -\f{\be}{2}+ 
 \f{1}{2} \sqrt{\f{\be^2+4}{2c_p+1}}
$$
Intersecting with the solutions of \eqref{110}, we obtain
\begin{theorem}
\label{theo:50}
Let $\vp_\om$ in \eqref{120} be the solutions to \eqref{100}, which exists for all $\om$ as in \eqref{110}.  These solutions of \eqref{10} 
are  spectrally stable  \underline{if and only if}
$$
\om \in \left(\f{-\be-\sqrt{\be^2+4}}{2},  -\f{\be}{2} - 
\f{1}{2} \sqrt{\f{\be^2+4}{2c_p+1}}\right] \cup \left[-\f{\be}{2} +
\f{1}{2} \sqrt{\f{\be^2+4}{2c_p+1}}, \f{-\be+\sqrt{\be^2+4}}{2}\right).
$$
\end{theorem}
Note that when $\be=0$, we arrive at the classical result that stability holds if and only if $p<1+\f{4}{d}$ and 
$
1>|\om|\geq \sqrt{\f{p-1}{4-(p-1)(d-1)}}.
$
\subsection{The case of $|\be|<<1$} 

In this section, we fix a decaying potential $W$, take a power nonlinearity $f(z)=z^{\f{p-1}{2}}$ and general $d \geq 1$.  We use $\be$ as a bifurcation parameter.  We first need to investigate the condition \eqref{cond:1}. 
We have 
$$
L_+(\be)=-\De+ (1-\om^2)-\be \om W- p \vp^{p-1}_{\om, \be}. 
$$
Our reference will be of course the operator $L_+(0)$, which is 
$
L_+(0)=L_+=-\De+ (1-\om^2)p \vp^{p-1}_{\om, 0}.
$
Here, $\vp^{p-1}_{\om, 0}=\vp_0$ is nothing but the unique even solution of $-\De \vp+(1-\om^2) \vp - \vp^p=0$. If we look for an asymptotic expansion for $\vp_{\om, \be}$, we should take it in the form 
$\vp_{\om, \be}= \vp_{0}+\be \psi_\om+O(\be^2)$. We have the defining  equation
$$
-\De \vp+(1-\om^2) \vp-\be \om W \vp - \vp^p=0.
$$
Taking a derivative in $\be$ yields 
$
L_+(\be)[\p_{\be} \vp]=\om W \vp. 
$
Setting $\be=0$ yields $L_+[\psi]=\om W \vp_0$.  Now, if $W\vp_0\perp Ker[L_0]=span[\p_1 \vp_0, \ldots, \p_d \vp_0]$ (this is certainly the case, if $W$ is an even function, which we assume henceforth) this last equation has  a solution, which is in the form 
$$
\vp_{\om, \be}=\vp_0+\be L_+^{-1}[W \vp_0]+O(\be^2). 
$$

With the formula $\vp_{\om, \be}=\vp_0+ \be \om L_+^{-1}[W \vp_0]+O(\be)$ in hand, let us now derive an asymptotic  formula for $L_+(\be)$ for $|\be|<<1$. We have 
\begin{eqnarray*}
L_+(\be) &=& -\De+(1-\om^2)-\be \om W - p(\vp_0+ \be \om L_+^{-1}[W \vp_0]+O(\be))^{p-1} \\
&=& L_+ -\be \om(W+p(p-1)L_+^{-1}[W \vp_0])+O(\be^2)
\end{eqnarray*}
How can we ensure that $L_+(\be)$ will have exactly one negative eigenvalue for small values of $\be$~? We are perturbing off $L_+$ which has exactly one negative eigenvalue and $d$ vectors in its kernel. By using the quadratic form characterization of the eigenvalues, we conclude that after the perturbation by $\be$, $L_+(\be)$ will still have at least this (perturbed) eigenvalue, but it is possible that some of the zero e-values will (after the perturbation) turn into negative ones as well. Clearly then, the criteria for $n(L_+(\be))=1$ for small $\be$ and $\om>0$  are in the form 
$$
\be(\dpr{W\p_j \vp_0}{\p_j \vp_0}+ p(p-1)\dpr{L_+^{-1}[W \vp_0] \p_j \vp_0}{\p_j \vp_0})<0, j=1, \ldots, d.
$$
At this point, we require that $W$ is radial (this in addition to the fact that $\vp_0$ is radial).  In this way, the expressions above are equal, for all values of $j=1, \ldots, d$. 
Thus, if we impose the non-degeneracy condition 
$$
\dpr{W\p_j \vp_0}{\p_j \vp_0}+ p(p-1)\dpr{L_+^{-1}[W \vp_0] \p_j \vp_0}{\p_j \vp_0}\neq 0
 $$
and let  $|\be|<<1$, so that  
$sgn(\be)=-sgn(\dpr{W\p_1 \vp_0}{\p_1 \vp_0}+ p(p-1)\dpr{L_+^{-1}[W \vp_0] \p_1 \vp_0}{\p_1 \vp_0})$ we will have ensured that $n(L_+(\be))=1$. This is because we have made sure that the zero eigenvalues bifurcated to become small positive eigenvalues.  
We can now state the perturbation result, whose proof we just outlined. 
\begin{theorem}
\label{theo:60}
Let $d\geq 1$, $p<1+\f{4}{d}$ and  $W=W(|x|)$  be a decaying   potential. Let $\vp_0$ be the unique radial solution of $-\De \vp+(1-\om^2)\vp-\vp^p=0$.  Assume  the following non-degeneracy condition holds 
for some interval $\om\in I\subset (0,1)$: 
\begin{equation}
\label{150}
\int_{\rd} (W+ p(p-1) L_+^{-1}[W \vp_0]) |\p_1 \vp_0(x)|^2 dx\neq 0\ \ \textup{for}\ \ \om\in I
\end{equation}
Then, there exists $\be_0: 0<\be_0<<1$, so that for all $\be: |\be|<\be_0$  and 
$$
sgn(\be)=-sgn(\int_{\rd} (W+ p(p-1) L_+^{-1}[W \vp_0]) |\p_1 \vp_0(x)|^2 dx)
$$
the Klein-Gordon equation \eqref{10} has a standing wave solution $e^{i \om t} \vp_{\om,\be}$. This solution is stable for $\om \in I$  if and only if 
\begin{equation}
\label{220}
\p_\om[\int(2\om+\be W) |\vp_{\om, \be}(x)|^2 dx]\leq 0, \om \in I.
\end{equation}
Moreover, the solution to the inequality  \eqref{220} is in the form 
$$
\om \in I\cap  (\om(p,d, \be),1),\ \ \  \om(p,d, \be)=\sqrt{\f{p-1}{4-(p-1)(d-1)}}+O(\be).
$$

\end{theorem}
{\bf Remark:} Note that if we choose $\be$ with the same sign as \\ $sgn(\int_{\rd} (W+ p(p-1) L_+^{-1}[W \vp_0]) |\p_1 \vp_0(x)|^2 dx)$, we will have  for   $|\be|<<1$  a total of $d+1$ negative eigenvalues for $L_+$, in which case Theorem \ref{theo:10} is inapplicable. We conjecture that generically we will   observe instability in this case.

\section{Numerical Results}
To conclude this brief contribution, we will test the main result of
the analysis, namely Theorem~\eqref{theo:20} through a numerical
case example.
In the numerical computation, we consider the discrete variant
of the model
\begin{equation}
\label{10-discrete}
\ddot{u}_{m}+ i \be W(x_m) \dot{u}_m- \frac{1}{(\Delta x)^2}(u_{m+1}-2u_m+u_{m-1})  +u_m - f(|u_m|^2) u_m=0 
\end{equation}
for a sufficiently small $\Delta x$ (typically $\Delta x=0.2$) is used such that the continuum model is well
approximated by $u_m(t) \equiv u(x_m,t)$, where $x_m=m \Delta x$.
We will present the stability results of the standing wave solutions in the form $e^{i\omega t}\varphi_m$ which satisfy
\begin{equation}
\label{20-discrete}
- \frac{1}{(\Delta x)^2}(\vp_{m+1}-2\vp_m+\vp_{m-1})  +(1-\om^2) \vp_m-\be \om W \vp_m-f(\vp_m^2)\vp_m=0.
\end{equation}
The numerical solution to this problem is identified via a Newton-type 
fixed point method. 

Once the relevant standing wave is obtained, following the prescription of
Section \ref{intro}, we linearize around it, obtaining a discrete analogue
of Eq.~\eqref{40}.  
An equivalent
formulation of this as a first order system reads:
$$
{\dot{\bf V}}_m=\mathcal{L}{\bf V}_m
$$
where
${\bf V}_m=\left(\begin{array}{c}{\bv}_m\\{\dot{\bv}}_m\end{array}\right)$ and
$
\mathcal{L}=\left(\begin{array}{cc}0 &1\\-\ch
&-J\end{array}\right).
$

\begin{figure}[tbp]
\begin{center}
\subfigure[]{{\includegraphics[width=7.75cm,height=7cm]{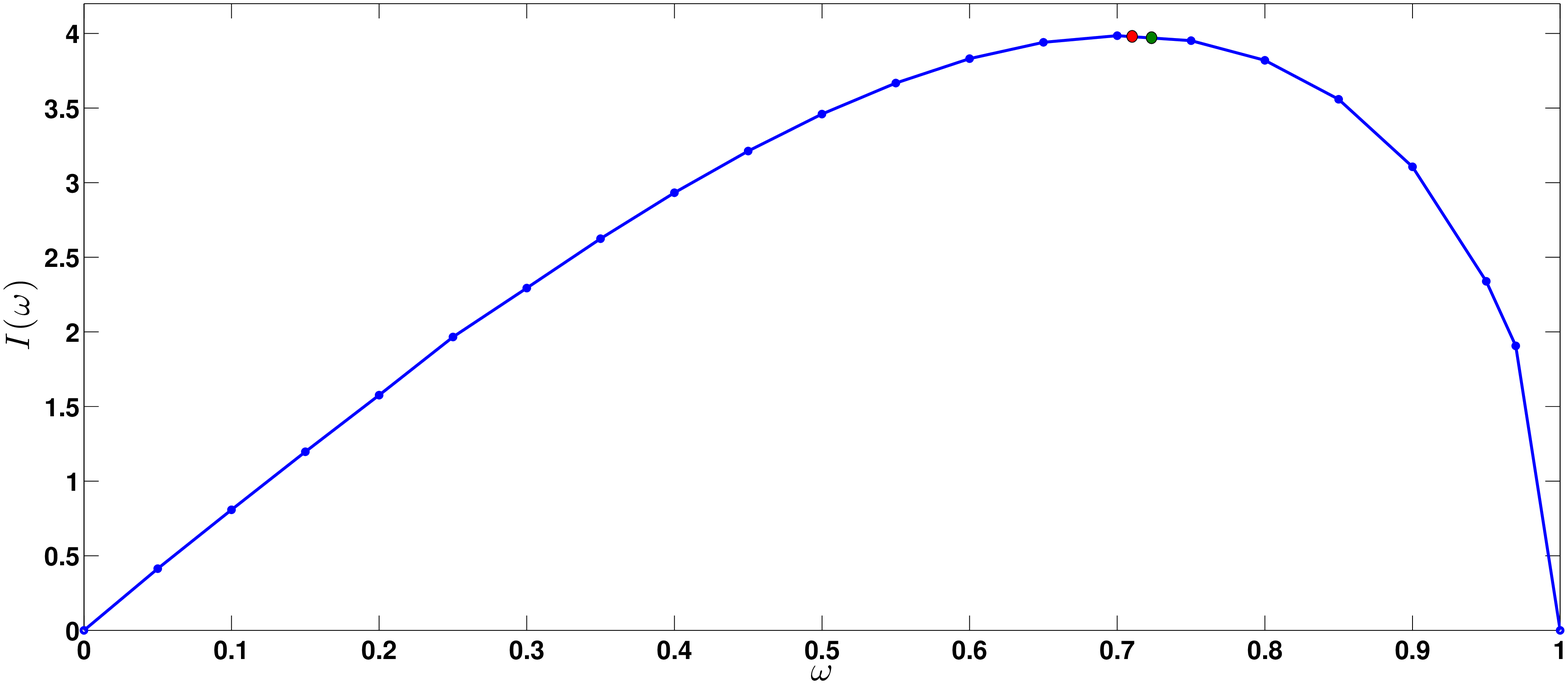}}}\\
\subfigure[]{{\includegraphics[width=7.75cm,height=5cm]{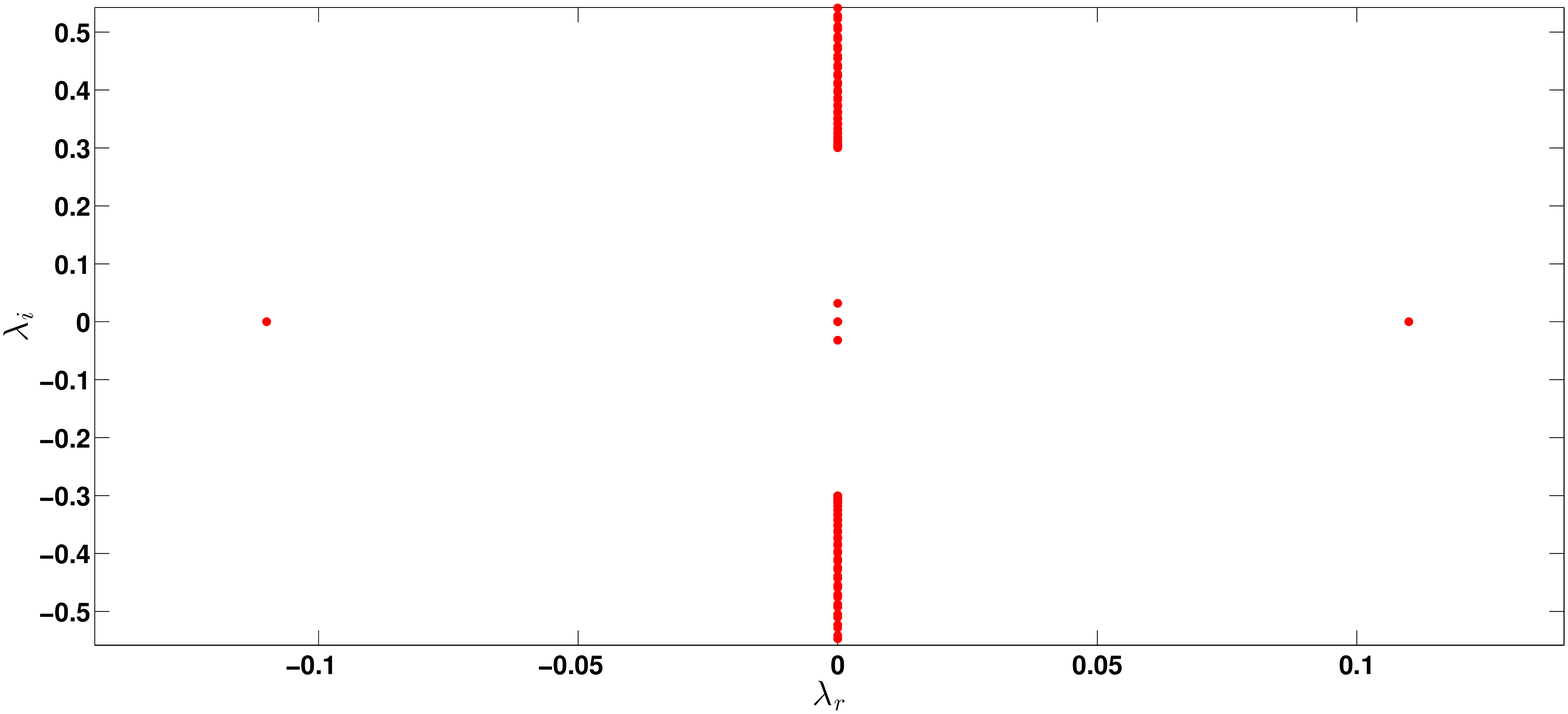}}}
\subfigure[]{{\includegraphics[width=7.75cm,height=5cm]{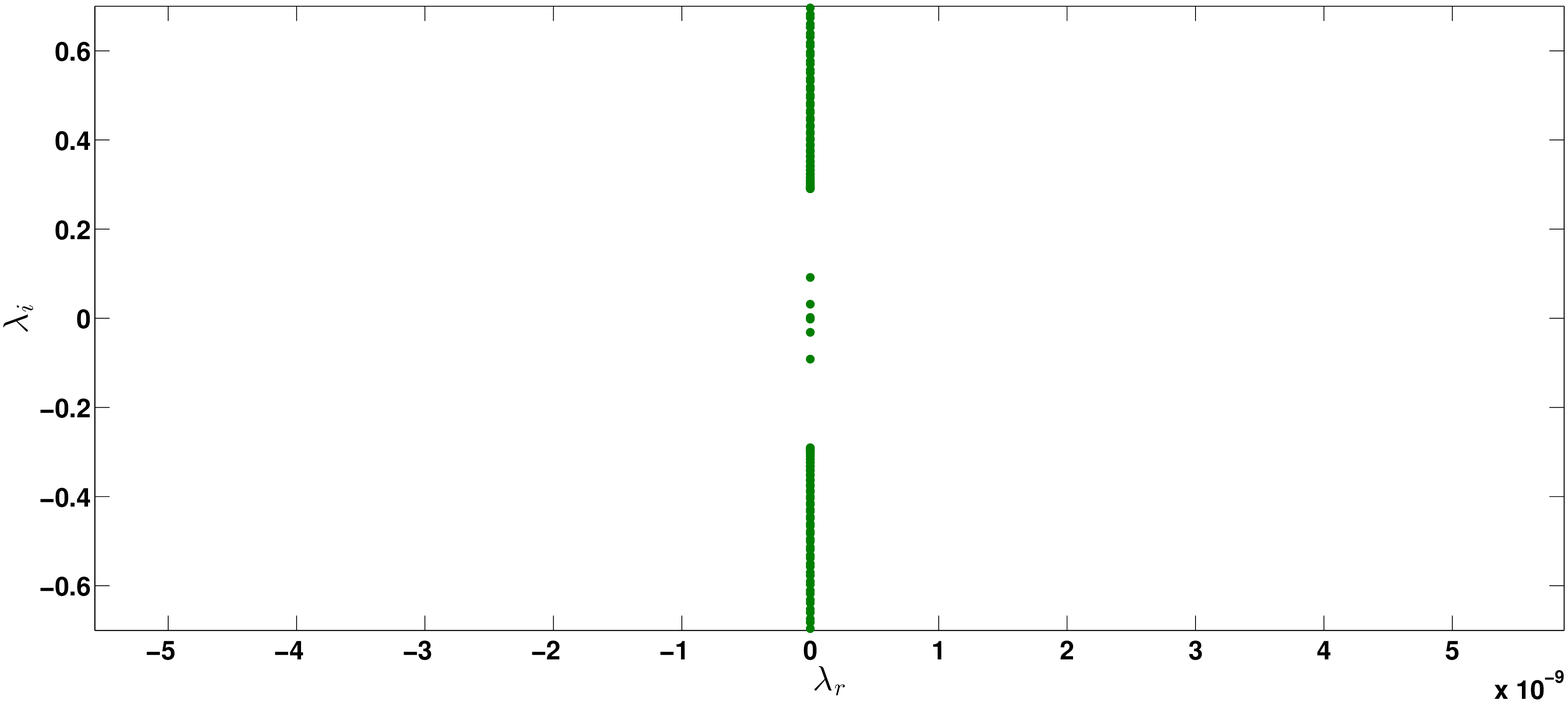}}}
\end{center}
\caption{(a) shows the relation between $\omega$ and $I(\omega)$ when $\beta=0.06$. The red circle on the graph corresponds to $\omega=0.7$ and the green one corresponds to $\omega=0.71$. (b) shows the spectral plane $(\lambda_r,\lambda_i)$ of the eigenvalues $\lambda=\lambda_r+i\lambda_i$ of $\mathcal{L}$ when $\omega=0.7$ and (c) shows the spectral plane $(\lambda_r,\lambda_i)$ of the eigenvalues $\lambda=\lambda_r+i\lambda_i$ of $\mathcal{L}$ when $\omega=0.71$. }
\label{FIG_1}
\end{figure}

As a case example of the even function $W(x)$, 
we use $W(x)=ae^{-bx^2}$ while $f(z)=z$ (i.e., we explore the
cubic nonlinearity) and $d=1$. A typical result of the numerical
computations is illustrated in
Figure \ref{FIG_1} (for the case of $\beta=0.06$, i.e.,
for small values of $\beta$). In panel (a), we plot the function 
$I(\om)= \int (2\om+ \be W)\vp_{n,\om}^2 $ for $\om \in [0,1]$, $a=0.1$, $b=1$. 
Based on the analytical prediction of Theorem~\eqref{theo:20}, 
there exists $\om ^{*} \in (0.705, 0.706)$ such that $I(\om)$ increases on $(0, \om ^{*})$ and decreases on $(\om ^{*},1)$. Accordingly, the theorem
predicts a {\it change of stability} as $\om ^{*}$ is crossed.
In order to see the spectral picture, we pick two $\om$ values, one slightly 
below $\om^{*}$ ($\om=0.7$) and one slightly above $\om^{*}$ ($\om=0.71$). Figure \ref{FIG_1} (b) clearly shows that there exists a real eigenvalue pair
of $\mathcal{L}$ when $\om=0.7$. This holds true for any $\om<\om^{*}$. However in Figure \ref{FIG_1} (c) we see that all the eigenvalues lie on the imaginary axis when $\om=0.71$, i.e., the standing wave is now neutrally 
stable, as theoretically predicted for $\om>\om^{*}$. Numerical results also 
show that as $\be$ increases, the value of $\om ^{*}$ decreases. 
Hence, our numerical computations indeed illustrate the sharpness
of the relevant criterion.


\section{Conclusions and Future Challenges}

In the present work, we have considered, motivated
by $\mathcal{PT}$-symmetric considerations and 
perturbations respecting the parity and time-reversal
symmetries, to explore a variant of
the recently explored $\mathcal{PT}$-symmetric Klein-Gordon
systems, which also can be thought of as a Schr{\"o}dinger
type equation with additional inertial terms. For this PDE,
we have explored the general stability criterion of~\cite{SS}
in order to derive a more precise/specific stability condition.
The latter has the natural form of an extension to the 
well-known Vakhitov-Kolokolov criterion of nonlinear Schr{\"o}dinger
type models. The relevant inequality has not only been
theoretically proposed and directly computed in some simple
case examples (such as $W(x)=1$), but its sharpness has been
numerically corroborated in analytically intractable forms of
the relevant function.

Nevertheless, there are numerous extensions of the present 
setting that may still be relevant to explore. While, partially
also due to space restrictions, here we constrained ourselves
and our numerical study to the prototypical cubic case, it would
be interesting to explore more general (power or other e.g. saturable)
nonlinearities that may well impart additional instabilities, as well as 
provide settings where the setup of the theorem will not apply. Here,
the numerical computations may provide insights towards
suitable generalizations. Also, considering more complex forms
of $W(x)$ may provide multiple changes of monotonicity (of the relevant
quantity of the theorem), and it
would be interesting to seek the corresponding stability reversals.
Finally, in the case of one or more of these instabilities, it would
be useful to dynamically investigate the fate of the resulting 
dynamical evolutions. Such studies are presently in progress and will
be reported in future studies.

\end{document}